\documentclass[twocolumn,showpacs,preprintnumbers,amsmath,amssymb]{revtex4}\usepackage{graphicx}
\usepackage{amsmath}
\usepackage{color}

\begin{document}

\title{Effects of strategy-migration direction and noise in the evolutionary spatial prisoner's dilemma}
\author{Zhi-Xi Wu}\thanks{zhi-xi.wu@physics.umu.se}
\author{Petter Holme}\thanks{petter.holme@physics.umu.se}
\affiliation{Department of Physics, Ume{\aa}, University, 901 87
Ume\aa, Sweden}
\begin{abstract}

Spatial games are crucial for understanding patterns of
cooperation in nature (and to some extent society). They are known
to be more sensitive to local symmetries than e.g.\ spin models.
This paper concerns the evolution of the prisoner's dilemma game
on regular lattices with three different types of neighborhoods
--- the von Neumann-, Moore-, and kagome types. We investigate two
kinds of dynamics for the players to update their strategies (that
can be unconditional cooperator or defector). Depending on the
payoff difference, an individual can adopt the strategy of a
random neighbor (a voter-model-like dynamics, VMLD), or impose its
strategy on a random neighbor, i.e., invasion-like dynamics
(IPLD). In particular, we focus on the effects of noise, in
combination with the strategy dynamics, on the evolution of
cooperation. We find that VMLD, compared to IPLD, better supports
the spreading and sustaining of cooperation. We see that noise has
nontrivial effects on the evolution of cooperation: maximum
cooperation density can be realized either at a medium noise
level, in the limit of zero noise, or in both these regions. The
temptation to defect and the local interaction structure determine
the outcome. Especially, in the low noise limit, the local
interaction plays a crucial role in determining the fate of
cooperators. We elucidate these both by numerical simulations and
mean-field cluster approximation methods.

\end{abstract}
 \maketitle

\section{Introduction}

In the last decades, a great deal of interest has been devoted to
understanding the evolution of cooperation~\cite{Axelrod}, i.e.,
how cooperation emerges and persists in a population composed of
selfish individuals. Evolutionary game
theory~\cite{Smith1982book,Weibull1995book} --- a research field
cultivated traditionally by both evolutionary biologists,
political scientists, economists  and
mathematician~\cite{Hofbauer1998book,Nowakbook} that recently has
attracted a growing interest from the physics
community~\cite{Szaboreview,Hauert2005ajp} ---  provides a general
framework to address such issues. One famous metaphor for the
problem of cooperation is the prisoner's dilemma (PD)
game~\cite{rapoport,Deobeli2005el}, in which defection could
benefit the individual in the short term, whereas cooperation has
only potential benefits in a longer time perspective. In the
common mathematical formulation of PD with pairwise interaction,
each of two encountering cooperators (defectors) get a payoff $R$
($P$), a defector confronting a cooperator acquires payoff $T$,
while the cooperator gains $S$. The four parameters is required to
satisfy the conditions $T>R>P>S$ and $2R>T+S$. In a well-mixed
population, defectors are unbeatable and cooperators are doomed to
extinction~\cite{Hofbauer1998book}.

The seminal work of Nowak and May~\cite{Nowak1992nature}, where a
PD game is played on two-dimensional grids with agents lacking
memory and ability to form complex strategies, showed that spatial
structure and nearest neighbor interactions can enable cooperators
to survive by forming clusters within which they benefit from
mutual cooperation and protecting them from exploitation by
defectors. This work has inspired numerous subsequent
investigations of evolutionary games on spatial
grids~\cite{Lindgren1994pd,Hauert2004nature}.
% It has also been
%incorporated into the  recent boom of network
%studies~\cite{Albert2002rmp}, where the evolutionary PD game has
%been studied on various model networks and real social
%networks~\cite{Abramson2001pre, Santos2005prl}, focusing on the
%structural effects (such as small-world property, average degree,
%degree heterogeneity, degree correlation, clustering, and
%hierarchical structure, etc) on the evolution of cooperation.
Now the spatial (or network) reciprocity is regarded as one of the
five main mechanisms supporting
cooperation~\cite{Nowak2006science,Nowakbook}. Furthermore,
coevolution of strategy distribution and underlying interaction
graphs can also be useful in characterizing the PD
game~\cite{Ebel2002pre,Zimmermann2004pre,Pacheco2006prl,Li2007pre}.
The crucial observation is that through switching partners,
cooperators are capable of frequently meeting other cooperators,
which substantially increase their reproduction rate. In addition,
a lots of other mechanisms favoring cooperation have also been
proposed: volunteering participation~\cite{Szabo2002prl},
separation of interaction and learning
graphs~\cite{Ohtsuki2007prl}, dilution and random diffusion of
agents on the grid~\cite{Vainstein2001pre}, success-driven
migration~\cite{Helbing2009pnas}, memory
effects~\cite{Wang2006pre}, reputation-based
interaction~\cite{Fu2008pre}, moderate aspiration
level~\cite{Alonso2006jsm}, etc. For a comprehensive review of
this research field, we refer to
Refs.~\cite{Szaboreview,Nowakbook}.

It is worth pointing out that the scale-free
topology, thanks to the capability of mutual
protection of the hubs if occupied by cooperators, can enhance the
spreading of cooperative behavior and resist the invasion of
defection, hence becoming a promoter of
cooperation~\cite{Santos2005prl}. However, it was also argued that
the adoption of averaged payoffs instead of accumulated ones in
the score function, as well as an introduction of participation
costs, might weaken this mechanism~\cite{Santos2006jeb}. These
results suggest that the distinct ability of strategy breeding of
the players~\cite{Szolnoki2008pa}, rather than the interaction
graph, matters much for the emergence of cooperation. The
heterogeneous or asymmetric strategy migration ability can be
implemented in different ways, such as dynamic preferential
selection~\cite{Wu2006pre}, introduction of two types of players
with different teaching activity~\cite{Szolnoki2007epl}, social
diversity~\cite{Perc2008pre,Szabo2009pre}, nonlinear attractive
effect~\cite{Guan2006epl,Chen2008pre}, etc. In principle, as long
as some distinguished players have higher influence to spread
their strategy~\cite{Szabo2009pre}, and also the connections among
these influential players are coupled in some appropriate
way~\cite{Szolnoki2008pa,Szolnoki2007epl,Perc2008pre}, the dilemma
can be relaxed and the promotion of cooperation is warranted.

These results suggest that the heterogeneity in the migration of
strategies might be an important factor for the stability of
cooperation. To our knowledge, in most previous studies of spatial
games, a common simplifying assumption is that, whenever updating
strategy, a player selects a neighbor as reference and attempts to
adopt the neighbor's strategy according to some prescribed
criterion. In other words, the focal player is always a recipient
of the strategy. Only few works have studied the reverse
situation, with strategy donors and their influence on the
evolution of cooperation. Ohtsuki \emph{et al.}\ implemented both
death-birth and birth-death updating for games on graphs, and
found that for birth-death updating, selection does not favor
cooperation~\cite{Ohtsuki2006nature}. Antal \emph{et al.}\ studied
evolutionary dynamics on degree-heterogeneous graphs, evolved
either by one individual dying and being replaced by the offspring
of a random neighbor (voter model dynamics) or by an individual
giving birth to an offspring taking over a random neighbor
(invasion process dynamics). They found that the fixation
probability of a single fitter mutant under the voter model
dynamics is $k^2$ times of corresponding value for the invasion
process dynamics ($k$ is the node degree)~\cite{Antal2006prl}.
With this paper, we continue investigating the effects of the
direction of strategy migration on the evolution of cooperation in
the spatially explicit PD game. Another theme of recent research
is the effects of noise on dynamic processes~\cite{noise}. Indeed,
noise (or mutation) plays an important role in the outcome of
evolutionary games~\cite{Foster1990tpb,Miekisz2008review}. In
recent years, the effects of noise on spatial games have also been
studied~\cite{Szabo1998pre,Szabo2005pre,
Ren2007pre,Guan2007pre,Perc2006njp}. The spatial PD model proposed
in~\cite{Nowak1992nature} is a deterministic cellular automaton
and can be extended to a stochastic version by introduction of
noise in different ways. Perhaps the first spatial PD model with a
stochastic strategy adoption process was proposed in
Ref.~\cite{nowak:noise}. We take the approach of  Szab\'o and
T\H{o}ke~\cite{Szabo1998pre}, where the players updates their
strategies according to a Fermi function and in which noise have
the role of temperature in the kinetic Ising model. For this
simple ``noise-guided'' evolutionary model, the authors found rich
dynamic phenomena --- two absorbing states consisting of only
cooperators and defectors separated by a coexistence region.
Particularly, the phase transition involving the extinction of
cooperators or defectors is found to belong to the universality
class of directed percolation~\cite{Szabo1998pre}. In a similar
vein, they also studied how the noise affects the phase diagram of
PD on different two-dimensional lattices and Newman--Watts
networks~\cite{Szabo2005pre}. In somewhat different approach, Perc
\emph{et al.}\ and Tanimoto introduced noise to the payoff
matrix~\cite{Perc2006njp,Tanimote2007pre}. The reported results
show a coherence-resonance phenomenon where the fraction of
cooperators reaches its maximum at an intermediate noise
level~\cite{Perc2006njp}. In Ref.~\cite{Szabo2005pre} the noise
effect in PD games was studied, mainly focusing on the threshold
of extinction of cooperation as a function of noise intensity. A
clear picture for how the average fraction of cooperation evolves
by varying the noise level under different intensities of the
temptation to defection, to the best of our knowledge, is still
lacking. In the present work, we will follow the research line of
Refs.~\cite{Antal2006prl,Ohtsuki2006nature,Szabo2005pre} to study
how the strategy migration direction and selection noise affect
the cooperation. In the rest of the paper, we first define our
model and then treat the problems sketched above both by computer
simulations and cluster approximation methods. Finally, we make
some discussions and draw our conclusion.

\section{Model}

In the present study, we consider the evolutionary PD on three
types of regular lattices with periodic boundary conditions,
namely square lattice with von Neumann neighborhood (von Neumann
lattice), square lattice with Moore neighborhood (Moore lattice),
and kagome lattice, such that the number of encounters of each
player are four, eight, and six, respectively. The reason we
concentrate on these simple lattices is twofold: first, previous
studies have found that local interaction does affect the
spreading of cooperation~\cite{Szabo2005pre, Hauert2004nature};
second, their regular structures allow us to implement mean-field
cluster approximation
analysis~\cite{Szaboreview,Hauert2005ajp,Szabo2005pre}.

Following many
studies~\cite{Szaboreview,Szabo1998pre,Szabo2005pre,Szabo2009pre},
we use the Nowak--May parameterization~\cite{Nowak1992nature} of
the spatial PD, i.e., the temptation to defect $T=b$ (where
$1<b<2$), the reward for mutual cooperation $R=1$, the punishment
of mutual defection $P=0$, and the sucker's payoff $S=0$.
Initially, we let the players be either unconditional cooperators
(C) or defectors (D) with equal probability. The evolution of
strategies is governed by random sequential updating. For each of
the considered lattice topologies, we first let a randomly chosen
player $i$ reap its payoff $P_i$ by playing the PD game with its
nearest neighbors. Then, we select a random neighbor $j$ of $i$,
and let it acquire its payoff $P_j$ by playing the game with its
nearest neighbors.

When updating strategy, the focal player $i$ can be of the role of
either strategy donor or strategy-recipient. If $i$ is the
strategy-recipient, we implement a voter-model-like dynamics
(VMLD)~\cite{Antal2006prl}, which is probably the most common
approach~\cite{Szaboreview,Szabo1998pre,Szabo2005pre,
Szabo2009pre,Perc2008pre,Perc2006njp,
Szolnoki2007epl,Szolnoki2008pa,Santos2005prl,
Ren2007pre,Guan2006epl,Guan2007pre}. In this case, the player $i$
imitates the strategy of $j$ with a probability dependent on their
payoff difference \cite{Szabo1998pre}
\begin{equation}
W_{ij}=\frac{1}{1+\exp{[(P_i-P_j)/\kappa]}}\label{eq1},
\end{equation}
where $\kappa\in[0,\infty)$ denotes the noise level (or, in the
language of evolutionary biology, selection
intensity~\cite{Nowak2006science}). In the $\kappa=0$ limit, the
adoption of a \emph{successful} strategy is deterministic, while
in the $\kappa\to\infty$ limit, the strategy learning is
blind~\cite{Szabo2005pre}. If, however, we treat the player $i$ as
the strategy donor, we implement exactly invasion-process-like
dynamics (IPLD) for the game~\cite{Antal2006prl}. In this case,
the neighbor $j$ will try to take the focal player $i$'s strategy
with a probability defined as
\begin{equation}
W_{ji}=\frac{1}{1+\exp{[(P_j-P_i)/\kappa]}}\label{eq2}.
\end{equation}
By tuning the values of $b$ and $\kappa$ in the framework of the
two updating schemes, we can obtain pictures of
how the noise intensity and direction of strategy migration impact the
final outcome of the evolutionary spatial PD game.

In our simulation, we consider the possible strategy-migration for
the players one by one according to a random sequence. One pass
through all the agents is called a \textit{Monte Carlo (MC)
sweep}. Between each MC sweep we reshuffle the sampling sequence.
The total population is of size $200\times200$ for the von Neumann
and Moore lattices, and $3\times100\times100$ for the kagome
lattice. When repeating the above described elementary steps the
system develops into a final stationary state characterized by the
average fraction of cooperators $f_c$, which is measured for the
last $3000$ sweeps of the total $2\times 10^4$. All the simulation
results presented below are averages over $20$ independent
realizations of initial strategy distribution.

\section{Results and discussion}

\begin{figure}
\includegraphics[width=\linewidth]{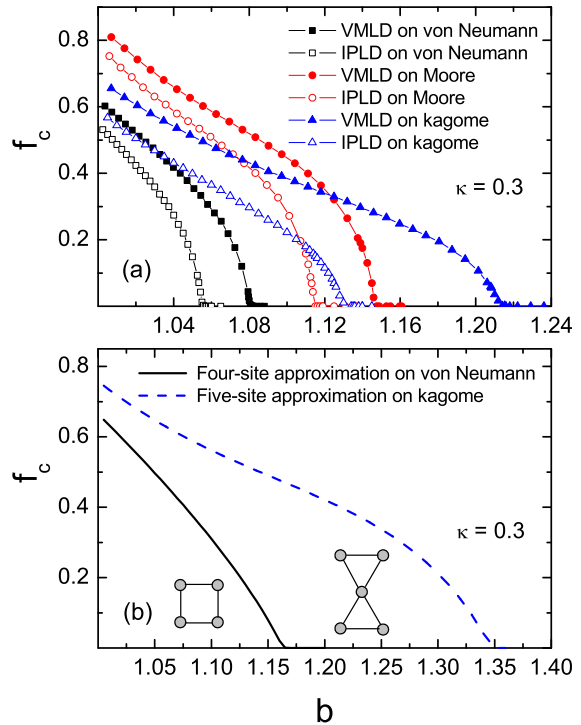}
\caption{(Color online) (a) Average fraction $f_c$ of cooperation
as a function of the temptation to defect $b$ for a fixed noise
intensity $\kappa=0.3$ on three types of lattices. Solid and open
symbols correspond to the two strategy-migration dynamics VMLD and
IPLD, respectively. (b) Theoretical estimations by the four-site
cluster approximation on the von Neumann lattice, and by the
five-site cluster approximation on the kagome lattice, correctly
predict the evolving tendency of $f_c$, but do not differentiate
between VMLD and IPLD dynamics. Note the different scaling of the
x-axis in the two panels. \label{fig1}}
\end{figure}

We start by comparing the results from the game with the VMLD
strategy-updating rule with the corresponding IPLD results. In
Fig.~\ref{fig1}(a), we present the simulation results for $f_c$ as
a function of the temptation to defect $b$ on the three lattices,
and the noise level $\kappa$ is fixed as $0.3$. We note that $f_c$
decreases monotonously with increasing $b$ up to a threshold
$b_c$, where cooperation vanishes. Although the qualitative
properties of the curves are somewhat similar, there exist
quantitative differences. By comparison, the stationary
cooperation level in IPLD is lower than the corresponding value
for VMLD. Furthermore, IPLD results in a lower $b_c$-threshold.
Thus, our first finding is that VMLD is more favorable for the
spreading of cooperation than IPLD.

\begin{figure}
\includegraphics[width=\linewidth]{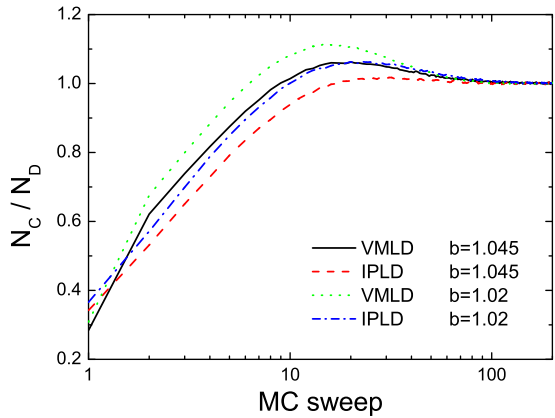} \caption{(Color
online) Time series of $N_C/N_D$, the number of events of $D$
switching to $C$ divided by the reverse case, for $b=1.02$ and
$1.045$ on the von Neumann lattice. \label{fig2}}
\end{figure}

At a first glance, it may seem strange that VMLD and IPLD on
homogeneous graph give rise to different results, since what is
regarded as VMLD from the point of view of the strategy-recipient
is exactly the IPLD in the viewpoint of the strategy donor. (On
heterogeneous graph, this difference is evident, since high-degree
individuals have greater chance to affect others for VMLD, and to
be affected by others for IPLD.) In fact, the four- or five-site
cluster approximation, developed in the spirit of mean-field
theory~\cite{Szaboreview}, do not differentiate between the VMLD
and IPLD, and give out exactly the same results for both strategy
updating dynamics (see Fig. 1(b), the two curves for the VMLD and
IPLD coincide with one another, and for simplicity we just show
one). A crucial observation is that this method neglects spatial
correlations in the population, which cause the update
probabilities of VMLD and IPLD to differ. Indeed, if the two
probabilities are the same, the final results will be identical on
homogeneous graphs. This argument can be deduced from the studies
of voter model on graphs. In Ref.~\cite{Castellano2005aip},
Castellano studied both voter and anti-voter dynamics on networks
(corresponding to the VMLD and IPLD in this work), and found that
on homogeneous network, the consensus time in both dynamics is the
same. In~\cite{Antal2006prl}, Antal \emph{et al.}\ pointed out
that the fixation probability of a neutral genotype on homogeneous
graph is the same for both voter-model and invasion-process
dynamics. The essential difference between the voter model and
evolutionary games is that in voter model, when a randomly
selected individual is to update its state, only the state of the
selected neighbor matters, while in the PD game case, both the
state of the selected neighbor and those of this neighbor's
neighbors. For our setup, the average payoffs collected by
cooperators on the boundary is greater than that by
defectors~\cite{Chen2008pre}, we argue that VMLD favors the
diffusion of cooperation on the rough boundaries separating
cooperators from defectors stronger than IPLD. (In the Appendix,
we have presented a concrete example to analyze the difference
between the VMLD and IPLD determining the probability of
strategy-migration.) A defector at a rough boundary would have, on
average, greater chance to be convinced by its cooperating
neighbors in the VMLD picture, than a cooperating neighbor to
convince a defector in the IPLD. To test this point, we have
traced the time series of $N_C/N_D$, the number of events of $D$
switching to $C$ divided by the reverse case, for two special
values of $b$ on the von Neumann lattice. The results presented in
Fig.~\ref{fig2} give further evidence for the above speculation.

\begin{figure}
\includegraphics[width=\linewidth]{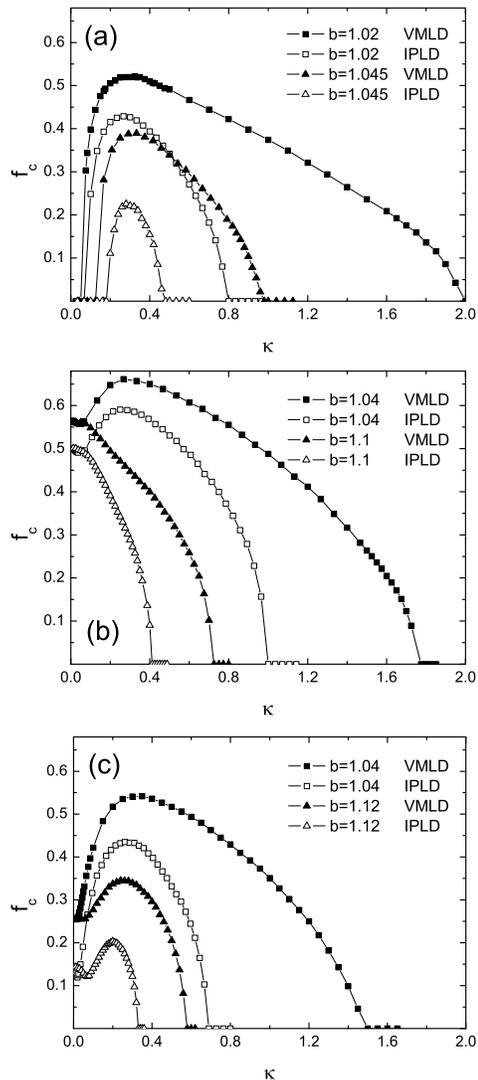} \caption{Average
fraction $f_c$ of cooperation as a function of the noise parameter
$\kappa$ for fixed values of $b$ on (a) the von Neumann lattice,
(b) the Moore lattice, and (c) the kagome lattice. The solid and
open symbols are as shown in Fig.~\ref{fig1}. The lines are guides
to the eye. \label{fig3}}
\end{figure}

Before moving forward, we briefly comment the importance of the
network structures in maintaining cooperation. In previous
studies~\cite{Szabo2005pre} the overlapping triangles (especially,
one-site overlapping triangles) in the connectivity structure are
found to support the spreading (maintenance) of cooperation. Here,
we want to point out that this conclusion is drawn in terms of the
magnitude of $b_c$ for the evolutionary PD on different lattice
structures~\cite{Szabo2005pre}. Looking back on our simulations in
Fig.~\ref{fig1}, we believe that this is not the whole story. If
we measure the capability of promoting cooperation by the average
fraction of cooperation in the equilibrium state, we observe that
when $b$ is small (e.g.\ $b=1.1$), the one-site overlapping
triangles (the kagome lattice) is an inferior structure compared
to the case of multiple overlapping sites (the Moore lattice).
This indicates that the temptation parameter is an indicative
variable when evaluating the role of connectivity structure,
giving a more nuanced picture than the previously mentioned effect
of overlapping triangles in Ref.~\cite{Szabo2005pre}.

\begin{figure}
\includegraphics[width=\linewidth]{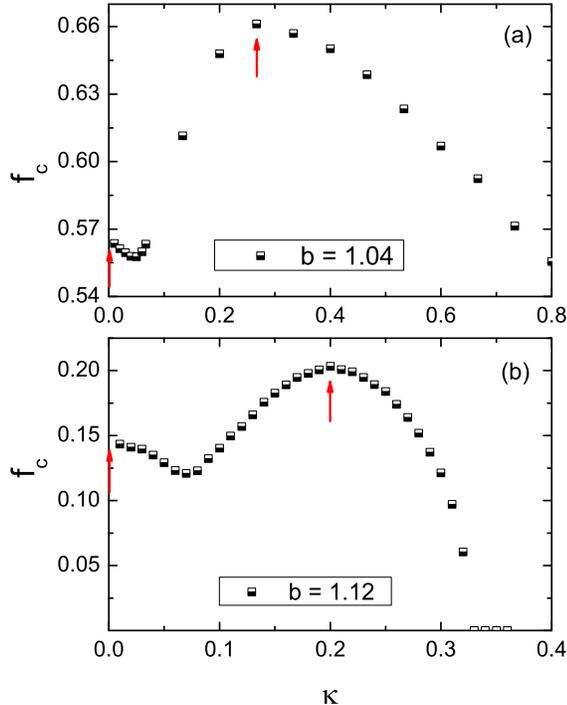} \caption{(Color online) Replotting of
the data in Fig.~\ref{fig3} for $b=1.04$ on the Moore lattice with
the VMLD (a), and $b=1.12$ on the kagome lattice with the IPLD
(b). The places where $f_c$ reaches its maximum value are
indicated by arrows.\label{fig4}}
\end{figure}

\begin{figure}
\includegraphics[width=\linewidth]{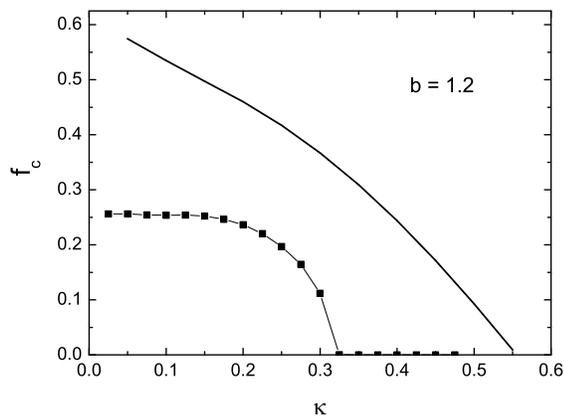} \caption{Average
fraction $f_c$ of cooperation versus $\kappa$ in the case of
``high'' temptation to defect $b=1.2$ on the kagome lattice. The
solid line denotes the analytical results obtained by the
five-site cluster approximation.\label{fig5}}
\end{figure}

Now let us turn to the effect of noise. As mentioned above, the
extinction threshold of cooperation $b_c$ as a function of
$\kappa$ on different types of lattices has been studied
in~\cite{Szolnoki2007epl,Szabo2005pre}. Here we investigate how
$\kappa$ influences $f_c$ in the stationary state for different
values of $b$. In Fig.~\ref{fig3}(a), we show simulation results
of $f_c$ versus $\kappa$ for two values of $b=1.02$, $b=1.045$ on
the von Neumann lattice. We observe so-called coherence resonance
for both VMLD and IPLD, where $f_c$ reaches its maximum at an
intermediate $\kappa$. If, in this case, the noise is too weak or
strong cooperation may be extinct (this is in accordance with
previous results~\cite{Perc2006njp}). For the parameters used, the
maximum $f_c$ emerges at about $\kappa=0.3$. We note that for the
same temptation to defect, comparing with the case of VMLD, IPLD
results in a narrower region of $\kappa$, where cooperators can
maintain a non-zero fraction in the population. That is to say,
cooperation is more robust against the fluctuation of noise in the
case of VMLD. This result strengthens our previous finding from
Fig.~\ref{fig1}. Note that $\lim_{\kappa\to\infty} f_c=1/2$. In this
paper we focus on low noise levels, i.e.\ situations where the competitive
interactions dominate the dynamics and do not study how cooperation
reemerges for very large $\kappa$.

\begin{figure}
\includegraphics[width=\linewidth]{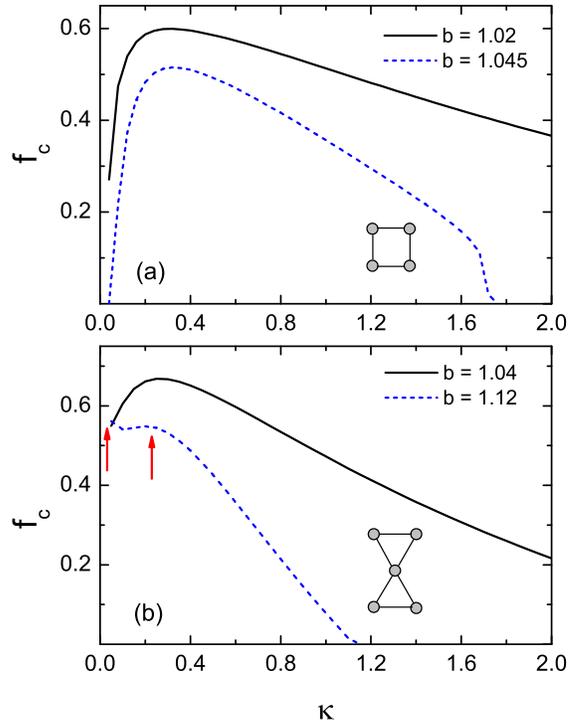} \caption{(Color online) Theoretical
estimations for $f_c$ as a function of $\kappa$ by using the
four-site cluster approximation method on the von Neumann lattice
(a), the five-site cluster approximation on the kagome lattice
(b). The arrows mark the places where $f_c$ maximizes when
$b=1.12$.\label{fig6}}
\end{figure}

The observed resonant behavior can be understood as follows. As
mentioned above, the noise parameter $\kappa$ measures the
stochastic uncertainties in the process of strategy learning. In
the limit of weak noise, the irrational choices of the individuals
become very rare. The strategies of those individuals who get
higher payoffs would always migrate successfully. Though the
payoffs of the defectors at the interface are on average smaller
than those of cooperators~\cite{Chen2008pre}, the highest payoff
is always obtained by those defectors surrounded by three other
cooperators. As such, if the individuals imitate with complete
rationality ($\kappa\to0$), the defective strategy would spread
more easily in the population, which does harm to the formation of
clusters of cooperators. On the other hand, in the limit of large
noise, the payoff information becomes less important in
determining the success of strategy migration. In such situation,
it would be definitely inefficient for the cooperators at the
interface to spread their behavior (though they acquire higher
payoff on average than those defectors~\cite{Chen2008pre}). If,
however, the noise $\kappa$ is appropriately chosen, i.e., not too
large or too small, the cooperators would have more chance to
diffusion, and the defectors with highest payoff would have less
probability to be followed by cooperators, leading to the
surviving and enhancement of cooperation. Henceforth the optimal
noisy intensity $\kappa$ emerges.

The experimental tests of the PD game on the Moore and kagome
lattices display somewhat similar results (Figs.~\ref{fig3}(b) and
(c)). For an appropriate temptation to defect, when varying
$\kappa$, we also notice an emergent of coherence resonance.
Moreover, the VMLD is also found to be superior to the IPLD in
promoting cooperation. Despite these two aspects, when $\kappa$ is
sufficiently small, the simulation results illustrated in
Fig.~\ref{fig3}(a) are in stark contrast to those in
Figs.~\ref{fig3}(b) and (c). In the former case, the cooperators
are doomed to extinction in the limit of zero noise, whereas they
can persist in the population with finite $f_c$ in the latter two
cases. We argue that the presence of overlapping
triangles~\cite{Szabo2005pre} in the latter two lattices
contributes to the difference, since triangles are absent in the
von Neumann lattice. Note, however, that the importance of
overlapping triangles is evaluated with $b_c$
in~\cite{Szabo2005pre}, and with $f_c$ here. Comparing the curves
for $b=1.04$ in Figs.~\ref{fig3}(b) and (c), keeping
Fig.~\ref{fig1} in mind, we conclude that for high temptation
values, one-site overlapping triangles gives more favorable
conditions for cooperation, than the multiple-site case. For low
temptation values the situation is reversed.

It is worth noting that the PD game on Moore and kagome lattices
there is a nontrivial dependence of $f_c$ on $\kappa$ (which
depends both on the temptation parameter and the direction of
strategy migration). In particular, for sufficiently small $b$, we
obtain only one maximum of $f_c$ as a function of $\kappa$ (the
curve in Fig.~\ref{fig3}(c) with $b=1.04$ and VMLD). With
increasing $b$, peculiarly, we observe two peaks of $f_c(\kappa)$,
one located at an intermediate value and the other at $\kappa=0$.
This phenomenon is yet more prominent in the case of IPLD (c.f.,
Fig.~\ref{fig4}). Finally, if the value of $b$ is sufficiently
large (the curves of $b=1.1$ in Fig.~\ref{fig3}(b) and also
Fig.~\ref{fig5}), the peak of $f_c$ at finite $\kappa$ disappears
and only the no-noise maximum remains. For comparison, in
Fig.~\ref{fig6} we present analytic results via the four-site
cluster approximation on the von Neumann lattice and five-site
cluster approximation on the kagome lattice (for details, please
see the Appendix in Ref.~\cite{Szaboreview}), and observe that the
theoretical estimations correctly predict the tendency of the
results obtained by our MC simulations.

\begin{figure}
\includegraphics[width=\linewidth]{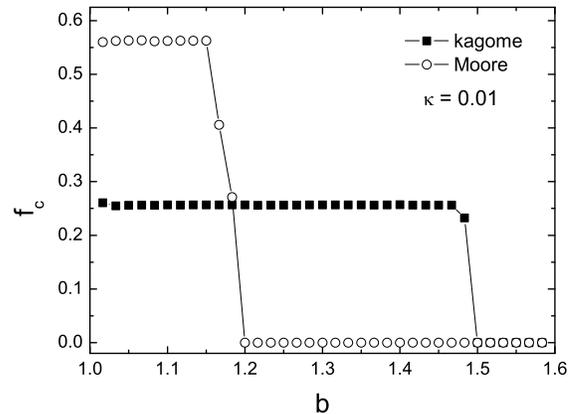} \caption{Average
fraction $f_c$ of cooperation as a function of $b$ in the case of
very weak noise $\kappa=0.01$ on the kagome lattice (solid
squares) and the Moore lattice (open circles).\label{fig7}}
\end{figure}

The simulation results shown in Figs.~\ref{fig3}, \ref{fig4}, and
\ref{fig5} allow us to speculate that in the limit of weak noise,
the local interaction topology may play a decisive role in
maintaining cooperation. To test this point further, we present in
Fig.~\ref{fig7} the simulation results for the evolutionary PD on
the Moore lattice and the kagome lattice, where $f_c$ as a
function of $b$ is plotted for a fixed value of $\kappa=0.01$.
Here we only consider the VMLD. We find that the cooperators
vanish at about $b_c=1.5$ for the kagome lattice (which has
already obtained in Ref.~\cite{Szabo2005pre}), and at about
$b_c=1.2$ for the Moore lattice. Despite the value of $b_c$, we
note that the average fraction of cooperation $f_c$ seems to be
insensitive to the temptation to defect staying as a plateau as
cooperators are capable of surviving (an effect not studied in
Ref.~\cite{Szabo2005pre}). In the following, we provide a local
configuration analysis for $f_c$ in the stationary state on kagome
lattices.

\begin{figure}
\includegraphics[width=0.8\linewidth]{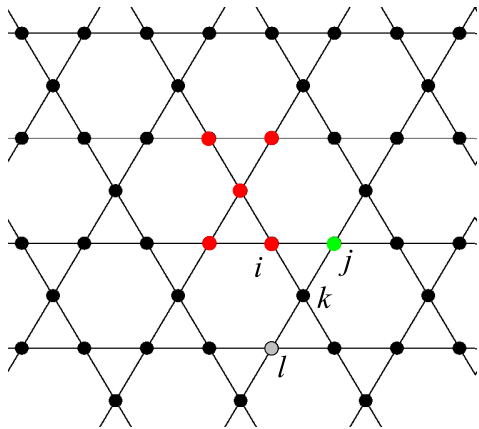} \caption{(Color online) Local
configuration stability analysis in the limit of zero noise on the
kagome lattice. The central five-site cluster of cooperators
cannot be destroyed as long as $b<3/2$ even if no neighbor or
next-nearest neighbor is a cooperator. See the text for
details.\label{fig8}}
\end{figure}

In the limit of weak noise, the strategy-updating rules
Eqs.~(\ref{eq1}) or (\ref{eq2}) is equivalent to the deterministic
imitation dynamics~\cite{Nowak1992nature}, i.e., as long as the
payoff of the strategy donor is greater than that of the strategy
recipient, the strategy will migrate. Since either an isolated
cooperator, or a connected pair of cooperators, cannot prevail
surrounded by defectors, consider a five-site cluster of
cooperators fully surrounded by defectors. If $b<3/2$, three
cooperators forming a triangle will expand its territory to a
five-site cluster, see Fig.~\ref{fig8}. In that case we may expect
that one of the neighboring defectors, say $j$, would since
($b<2$) imitate $i$'s strategy in the next sweep. Then their
common neighbor $k$ may acquire the payoff $2b$ in the next sweep.
However, since $i$ gets a total payoff $3$; then, as long as
$b<3/2$, there is no chance that $k$'s strategy can migrate to
$i$. In other words, such a local configuration in a sea of
defectors if $b<3/2$ is stable. However, if a neighbor $l$ of $k$
was a cooperator, the total payoff of $k$ would be $3b$ (which is
greater than $3$), and in that case $k$'s strategy could migrate
to $i$, hence destroying the original cluster. Taken together, the
necessary condition for a five-site cluster of cooperators
existing stably on the kagome lattice requires that $b<3/2$ and,
at the same time, there is no cooperator in their next-nearest
neighborhood. Since the second-nearest neighbors may be shared by
other five-site cluster of cooperators, we can estimate $f_c$ to
$5/(5+8+14/2)=1/4$ ($8$ and $14$ are the number of nearest and
next-nearest neighbors of the five-site cluster). This estimation
is very close to the one obtained by MC simulations $0.256(2)$.
Thus this local-configuration stability analysis gives a good
prediction for $f_c$ of evolutionary PD in the limit of weak noise
on the kagome lattice. Indeed, we have visualized the networked
population and found many isolated five-site clusters of
cooperators on the kagome lattice (results now shown here). The
same analysis, with the same conclusions can be done for the Moore
lattice.

The above results might shed some light on the stability of
cooperation in society. Real social networks often have distinct
communities, where individuals in one groups is densely connected,
having relatively few connections to other
groups~\cite{Girvan2002pnas}. The emergence of community structure
may hinder the spread of defection, just as in the case of the
five-site cluster on the kagome lattice above. In this sense, our
society may not be as sensitive to defection as indicated by the
PD on the von Neumann lattice. On the other hand, we also show
that VMLD is more efficient in maintaining and promoting
cooperation than IPLD. Therefore, in the social situations, if
everyone tries learn more from others, it would perhaps be easier
for people to work together efficiently.

\section{Conclusion}

In summary, we have studied the effects of the direction of
strategy migration and noise on the evolution of cooperation in
the spatial prisoner's dilemma game. To this end, we have
considered two types of strategy updating dynamics, namely
voter-model-like dynamics and invasion-process-like dynamics, on
three types of lattice structures --- square lattice with von
Neumann and Moore neighborhoods, and kagome lattice. It was found
that the VMLD, rather than the IPLD, better sustain and promote
cooperation on all the tested lattices. Furthermore, we found
noise to have a nontrivial effect on the evolution of cooperation
in spatial population. First, for low temptation to defect,
coherence resonance is found for all three types of lattice, i.e.,
there is an optimal noise level for promoting cooperation. Second,
for an intermediate temptation to defect, we observe a two-peak
behavior (for Moore and kagome lattices) --- the first maxima at
zero noise, the other at a moderate noise levels (on the von
Neumann lattice, however, we observe no such behavior). Third, for
even higher temptation levels, only no noise maximizes
cooperation. Moreover, we find that in the low-noise limit, the
local lattice structure determines the fate of cooperators. We
find, in the case of high temptation values, a structure of
triangles overlapping at one site to benefit cooperation compared
to a structure of triangles overlapping at multiple sites. For low
temptation values, the situation is reversed and the multiply
overlapping triangles are beneficial. Our MC-simulation results
are in good agreement with theoretical predictions obtained from
mean-field cluster approximation methods. These results may enrich
our knowledge of the evolution of cooperation in spatially
structured populations. A possible extension of the present work
is to consider evolutionary PD on degree-heterogeneous networks
with the incorporation of asymmetric direction of strategy
migration. We expect that degree-heterogeneity would induce a
stronger difference in the final results, as was found in voter
model on such networks~\cite{Antal2006prl,Castellano2005aip}.

\begin{acknowledgments}
The authors are greatly indebted to Dr.\ Jeromos Vukov for
numerous discussions of the five-site cluster algorithm. This
research was supported by the Swedish Research Council and the
Swedish Foundation for Strategic Research.
\end{acknowledgments}

\appendix

\section{A concrete example}

\begin{figure}[h]
\includegraphics[width=\linewidth]{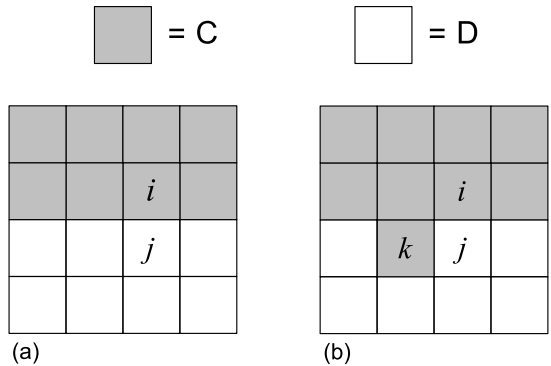} \caption{Illustration of
a small system with $4\times4$ players. (a) Both VMLD and IPLD
give out the same transformation probability for $j$ to change
from a defector to a cooperator. (b) The VMLD favors better the
diffusion of cooperation than the IPLD when the interface is
rugged.\label{fig9}}
\end{figure}

In this Appendix we present a small example illustrating how the
voter-like dynamics supports cooperation better than the
invasion-like dynamics. For the sake of simplicity we consider a
small system --- a $4\times4$ square lattice with von Neumann
neighborhood and periodic boundary conditions. Initially, half of
the sites of the lattice are occupied by cooperators, and the
other half by defectors as shown in Fig.~\ref{fig9}(a). We focus
our attention on the probability of strategy-migration of one
focal player, say $j$, changing from $D$ to $C$. In the case of
VMLD, this may happen when the individual $j$ is selected to
update its state and choose the neighbor $i$ as reference, which
happens with probability $1/4$. According to the payoff matrix of
the PD, the individual $i$ will get payoff $3$, since there are
three cooperators and one defector in its neighborhood. In
contrast, the individual $j$ acquires only payoff $b$, which comes
from the interaction with $i$. As a consequence, the expected
probability for the player $j$ changing from $D$ to $C$ is
\begin{equation} W_{D\to
C}=\frac{1}{4}\frac{1}{1+\exp[(b-3)/\kappa]}
\end{equation} if VMLD is adopted. This is
also the transformation probability if IPLD is implemented, rather
than VMLD. Thus, on a smooth interface, both VMLD and IPLD will
result in the same probability for strategy-migration. Since the
defectors on the interface gain much less payoff than their
cooperative neighbors ($b$ is just little larger than $1$), we can
expect cooperators could invade some of them as time evolves. Now,
assume the configuration in Fig.~\ref{fig9}(b), where the player
$j$ is surrounded by two cooperators, $i$ and $k$. In such case,
if $j$ is selected to update strategy and VMLD is used, the
probability for the player to become a cooperator is
\begin{align}W_{D\to C}&=
\frac{1}{4}\left(\frac{1}{1+\exp[(2b-3)/\kappa]}+
\frac{1}{1+\exp[(2b-1)/\kappa]}\right)\nonumber\\
&=\frac{1}{4}(w_{ij}+w_{kj})\label{VMLD}
\end{align}
where $w_{ij}/4$ and $w_{kj}/4$ denote, respectively, the
probability of $i$'s strategy, $k$'s strategy convincing $j$. If,
however, IPLD is adopted, the individual $j$ could be imposed on
the cooperative strategy either by $i$ or $k$, whose probability
is given by
\begin{align}W_{D\to
C}&=1-(1-w_{ij}/4)(1-w_{kj}/4)\nonumber\\  &=\frac{1}{4}
\left(w_{ij}+w_{kj}-\frac{w_{ij}w_{kj}}{4}\right),
\end{align}
which is smaller than the corresponding value from
Eq.~(\ref{VMLD}). Note that we have assumed the neighbors of $i$,
$j$, and $k$, have not changed their strategies between the two
events (from the updating time of $i$ to that of $j$). In sum this
illustrates how the VMLD favors the migration of cooperative
strategy along the interface separating cooperators and defectors
as compared to the IPLD. Note that this argument cannot be
reversed to hold for cooperators instead of defectors (considering
say $i$ in Fig.~\ref{fig9}(a) being a defector). The reason is
that, on average, the payoffs of defectors along the interface are
smaller than those of the
cooperators~\cite{Szabo2002prl,Chen2008pre}, which disables the
benefit of the VMLD, and thereby inhibits the diffusion of
defection.

\bibliographystyle{h-physrev3}

\begin{thebibliography}{99}
%\bibitem{Ball}
%P. Ball, Complexus \textbf{1}, 190 (2003).

%\bibitem{Castellano2007}
%C. Castellano, S. Fortunato, and V. Loreto, arXiv:0710.3256v1
%[Rev. Mod. Phys. (to be published)].

\bibitem{Axelrod}
R. Axelrod and W.D. Hamilton, Science \textbf{211}, 1390 (1981);
R. Axelrod, \emph{The Evolution of Cooperation} (Basic Books, New
York, Revised Edition, 2006).

\bibitem{Smith1982book}
J. Maynard Smith, \emph{Evolution and the theory of games}
(Cambridge University Press, Cambridge, 1982).

\bibitem{Weibull1995book}
J.W. Weibull, \emph{Evolutionary Game Theory} (MIT Press,
Cambridge, MA, 1995).

\bibitem{Hofbauer1998book}
J. Hofbauer and K. Sigmund, \emph{Evolutionary Games and
Population Dynamics} (Cambridge University Press, Cambridge,
1998).

\bibitem{Nowakbook}
M.A. Nowak, \emph{Evolutionary Dynamics: Exploring the Equations
of Life} (Harvard University Press, Cambridge, MA, 2006).

\bibitem{Szaboreview}%game review
G. Szab\'o and G. Fath, Phys. Rep. \textbf{446}, 97 (2007).

\bibitem{Hauert2005ajp}%game brief review
C. Hauert, G. Szab\'o, Am. J. Phys. \textbf{73}, 405 (2005).

\bibitem{rapoport}
A. Rapoport and A.M. Chammah, \emph{Prisoner's dilemma},
(University of Michigan Press, Ann Arbor, MI, 1965).

\bibitem{Deobeli2005el}%game brief review
M. Doebeli, C. Hauert, Ecol. Lett. \textbf{8}, 748 (2005).

\bibitem{Nowak1992nature}%lattice game
M.A. Nowak and R.M. May, Nature (London) \textbf{359}, 826 (1992);
Int. J. Bifurcat. Chaos \textbf{3}, 35 (1993).

\bibitem{Lindgren1994pd}
K. Lindgren and M.G. Nordahl, Physica D \textbf{75}, 292 (1994);
M. Nakamaru, H. Matsuda, and Y. Iwasa, J. Theor. Biol.
\textbf{184}, 65 (1997); F. Schweitzer, L. Behera, and H.
M\"uhlenbein, Adv. Complex Syst. \textbf{5}, 269 (2002).

\bibitem{Hauert2004nature}%PD and snowdrift
C. Hauert and M. Doebeli, Nature (London) \textbf{428}, 643
(2004).

%\bibitem{Albert2002rmp}
%R. Albert and A.-L. Barab\'asi, Rev. Mod. Phys. \textbf{74}, 47
%(2002); S.N. Dorogovtsev and J.F.F. Mendes, Adv. Phys.
%\textbf{51}, 1079 (2002); M.E.J. Newman, SIAM Rev. \textbf{45},
%167 (2003); S. Boccaletti, V. Latora, Y. Moreno, M. Chavez, and
%D.-U. Hwang, Phys. Rep. \textbf{424}, 175 (2006).

%\bibitem{Abramson2001pre}%networked game
%G. Abramson and M. Kuperman, Phys. Rev. E \textbf{63}, 030901(R)
%(2001); B.J. Kim, A. Trusina, P. Holme, P. Minnhagen, J.S. Chung,
%and M. Y. Choi, \emph{ibid.} \textbf{66}, 021907 (2002); P. Holme,
%A. Trusina, B.J. Kim, and P. Minnhagen, \emph{ibid.}
%\textbf{68}, 030901(R) (2003); J. Vukov and G. Szab\'o,
%\emph{ibid.} \textbf{71}, 036133 (2005); Z.-X. Wu, X.-J. Xu, Y.
%Chen, and Y.-H. Wang, \emph{ibid.} \textbf{71}, 037103 (2005);
%F.C. Santos, J.F. Rodrigues, and J.M. Pacheco, \emph{ibid.}
%\textbf{72}, 056128 (2005); M. Tomassini, L. Luthi, and M.
%Giacobini, \emph{ibid.} \textbf{73}, 016132 (2006); Z. Rong, X.
%Li, and X. Wang, \emph{ibid.} \textbf{76}, 027101 (2007); W.-X.
%Wang, J. Lu, G. Chen, and P.M. Hui, \emph{ibid.} \textbf{77},
%046109 (2008); S. Assenza, J.G\'omez-Garde\~nes, and V. Latora,
%\emph{ibid.} \textbf{78}, 017101 (2008); V. Hatzopoulos and H.J.
%Jensen, \emph{ibid.} \textbf{78}, 011904 (2008); S. Devlin and T.
%Treloar, \emph{ibid.} \textbf{79}, 016107 (2009); J.
%G\'omez-Garde\~nes, M. Campillo, L.M. Flor\'ia, and Y. Moreno,
%Phys. Rev. Lett. \textbf{98}, 108103 (2007); N. Masuda and K.
%Aihara, Phys. Lett. A \textbf{313}, 55 (2003); C.-L. Tang, W.-X.
%Wang, X. Wu, and B.-H. Wang, Eur. Phys. J. B \textbf{53}, 411
%(2006).

\bibitem{Santos2005prl}%scale-free
F.C. Santos and J.M. Pacheco, Phys. Rev. Lett. 95, 098104 (2005);
F.C. Santos, J.F. Rodrigues, and J.M. Pacheco, Proc. R. Soc.
London, Ser, B \textbf{273}, 51 (2006); F.C. Santos, J.M. Pacheco,
and T. Lenaerts, Proc. Natl. Acad. Sci. U.S.A. \textbf{103}, 3490
(2006).

\bibitem{Nowak2006science}%five rules
M.A. Nowak, Science \textbf{314}, 1560 (2006); M.A. Nowak and K.
Sigmund, In \emph{Theoretical Ecology: Principles and
Applications}, eds. R.M. May and A. McLean, Oxford: Oxford
University Press, Pages 7-16 (2007).

\bibitem{Ebel2002pre}%coevolution
H. Ebel and S. Bornholdt, Phys. Rev. E \textbf{66}, 056118 (2002).

\bibitem{Zimmermann2004pre}%coevolution solves dilemmas
M.G. Zimmermann, V.M. Egu\'iluz, and M. San Miguel, Phys. Rev. E
\textbf{69}, 065102(R) (2004); M.G. Zimmermann and V.M. Egu\'iluz,
\emph{ibid.} \textbf{72}, 056118 (2005); J. Tanimoto, \emph{ibid.}
\textbf{76}, 021126 (2007); F. Fu, T. Wu, and L. Wang,
\emph{ibid.} \textbf{79}, 036101 (2009).

\bibitem{Pacheco2006prl}%coevolution network and game dyanmics
J.M. Pacheco, A. Traulsen, and M.A. Nowak, Phys. Rev. Lett.
\textbf{97}, 258103 (2006); J. Theor. Biol. \textbf{243}, 437
(2006); F.C. Santos, J.M. Pacheco, and T. Lenaerts, PLOS Comput.
Biol. \textbf{2}, 1284 (2006); P. Holme and G. Ghoshal, Phys. Rev. Lett.
\textbf{96}, 098701 (2006).

\bibitem{Li2007pre}%coevolution network and game dyanmics
W. Li, X. Zhang, and G. Hu, Phys. Rev. E \textbf{76}, 045102(R)
(2007); R. Suzuki, M. Kato, and T. Arita, Phys. Rev. E
\textbf{77}, 021911 (2008).

\bibitem{Szabo2002prl}%volunteering participation
G. Szab\'o and C. Hauert, Phys. Rev. Lett. \textbf{89}, 118101
(2002); Phys. Rev. E \textbf{66}, 062903 (2002); C. Hauert and G.
Szab\'o, Complexity \textbf{8}, 31 (2003); G. Szab\'o and J.
Vukov, Phys. Rev. E \textbf{69}, 036107 (2004).

\bibitem{Ohtsuki2007prl}%separation of interaction and learning networks
H. Ohtsuki, M.A. Nowak, and J.M. Pacheco, Phys. Rev. Lett.
\textbf{98}, 108106 (2007); H. Ohtsuki, J.M. Pacheco, and M.A.
Nowak, J . Theor. Biol. \textbf{246}, 681 (2007); Z.-X. Wu and
Y.-H. Wang, Phys. Rev. E \textbf{75}, 041114 (2007); M. Zhang and
J. Yang, \emph{ibid.} \textbf{79}, 011121 (2009).

\bibitem{Vainstein2001pre}% dilute and random diffusion
M.H. Vainstein and J.J. Arenzon, Phys. Rev. E \textbf{64}, 051905
(2001); M.H. Vainstein, A.T.C. Silva, and J.J. Arenzon, J. Theor.
Biol. \textbf{244}, 722 (2007).

\bibitem{Helbing2009pnas}%success-driven migration
D. Helbing and W. Yu, Proc. Natl. Acad. Sci. U.S.A. \textbf{106},
3680 (2009).

\bibitem{Wang2006pre}%memory effect on snowdrift
W.-X. Wang, J. Ren, G. Chen, and B.-H. Wang, Phys. Rev. E
\textbf{74}, 056113 (2006); S.-M. Qin, Y. Chen, X.-Y. Zhao, and J.
Shi, Phys. Rev. E \textbf{78}, 041129 (2008).

\bibitem{Fu2008pre}%reputation
F. Fu, C. Hauert, M.A. Nowak, and L. Wang, Phys. Rev. E
\textbf{78}, 026117 (2008).

\bibitem{Alonso2006jsm}%aspiration level
J. Alonso, A. Fern\'andez, and H. Fort J. Stat. Mech. P06013
(2006); X. Chen and L. Wang, Phys. Rev. E \textbf{77}, 017103
(2008); T. Platkowski and P. Bujnowski, \emph{ibid.} \textbf{79},
036103 (2009).

\bibitem{Santos2006jeb}% scale-free advantage is hindered
F.C. Santos and J.M. Pacheco, J. Evol. Biol. \textbf{19}, 726
(2006); Z.-X. Wu, J.-Y. Guan, X.-J. Xu, and Y.-H. Wang, Physica A
\textbf{379}, 672 (2007); N. Masuda, Proc. R. Soc. London, Ser. B
\textbf{274}, 1815 (2007).

\bibitem{Szolnoki2008pa}%scale-free heterogeneous
A. Szolnoki, M. Perc, and Z. Danku, Physica A \textbf{387}, 2075
(2008); Europhys. Lett. \textbf{84}, 50007 (2008).

\bibitem{Wu2006pre}%heterogeneous imitation rate
Z.-X. Wu, X.-J. Xu, Z.-G. Huang, S.-J. Wang, and Y.-H.
Wang, Phys. Rev. E \textbf{74}, 021107 (2006);
Z.-X. Wu, X.-J. Xu, and Y.-H. Wang, Chin. Phys. Lett. \textbf{23},
531 (2006).

\bibitem{Szolnoki2007epl}%AB type
A. Szolnoki and G. Szab\'o, Europhys. Lett. \textbf{77}, 30004
(2007); A. Szolnoki, M.Perc, and G. Szab\'o, Eur. Phys. J. B
\textbf{61}, 505 (2008).

\bibitem{Perc2008pre}%social diversity and restricted connections
M. Perc, A. Szolnoki, and G. Szab\'o, Phys. Rev. E \textbf{78},
066101 (2008); A. Szolnoki and M. Perc, New J. Phys. \textbf{10},
043036 (2008); Eur. Phys. J. B \textbf{67}, 337 (2009).

\bibitem{Szabo2009pre}%increasing number of neighbors
G. Szab\'o and A. Szolnoki, Phys. Rev. E \textbf{79}, 016106
(2009).

\bibitem{Guan2006epl}%heterogeneous imitation
J.-Y. Guan, Z.-X. Wu, Z.-G. Huang, X.-J. Xu, and Y.-H. Wang,
Europhs. Lett. \textbf{76}, 1214 (2006).

\bibitem{Chen2008pre}%heterogeneous imitation
X. Chen, F. Fu, and L. Wang, Phys. Rev. E \textbf{78}, 051120
(2008).

\bibitem{Ohtsuki2006nature}% birth-death or death-birth
H. Ohtsuki, C. Hauert, E. Lieberman, and M. A. Nowak Nature
(London) \textbf{441}, 502 (2006).

\bibitem{Antal2006prl}%voter and invasion dyanmics
T. Antal, S. Redner, and V. Sood, Phys. Rev. Lett. \textbf{96},
188104 (2006); V. Sood, T. Antal, and S. Redner, Phys. Rev. E
\textbf{77}, 041121 (2008).

\bibitem{noise}
W. Horsthemke and R. Lefever, \emph{Noise-induced Transitions},
(Berlin: Springer-Verlag, 1984); J. Garc\'ia-Ojalvo and J.M.
Sancho, \emph{Noise in Spatially Extended Systems}, (New York:
Springer, 1999).

\bibitem{Foster1990tpb}
D. Foster and P. Young, Theor. Popul. Biol. \textbf{38}, 219
(1990); M. Kandori, G.~J. Mailath, R. Rob, Econometrica,
\textbf{61}, 29 (1993); L.~E. Blume, Games Econ. Behav.
\textbf{44}, 251 (2003).

\bibitem{Miekisz2008review}
J. Miekisz, in  \emph{Multiscale Problems in the Life Sciences},
edited by V. Capasso and M. Lachowicz (Springer, Berlin 2008).

\bibitem{Szabo1998pre}%introduction of noise
G. Szab\'o and C. T\H{o}ke, Phys. Rev. E \textbf{58}, 69 (1998).

\bibitem{Szabo2005pre}%noise phase diagrams
G. Szab\'o, J. Vukov, and A. Szolnoki, Phys. Rev. E \textbf{72},
047107 (2005); J. Vukov, G. Szab\'o, and A. Szolnoki, \emph{ibid.}
\textbf{73}, 067103 (2006); \emph{ibid.} \textbf{77}, 026109
(2008).

\bibitem{Ren2007pre}%noise in selction temperature
J. Ren, W.-X. Wang, and F. Qi, Phys. Rev. E \textbf{75}, 045101(R)
(2007).

\bibitem{Guan2007pre}%noise in selection temperature
J.-Y. Guan, Z.-X. Wu, and Y.-H. Wang, Phys. Rev. E \textbf{76},
056101 (2007).

\bibitem{Perc2006njp}%noise in payoff
M. Perc, New J. Phys. \textbf{8}, 22 (2006); M. Perc and M. Marhl,
\emph{ibid.} \textbf{8}, 142 (2006); M. Perc, \emph{ibid.}
\textbf{8} 183 (2006); Europhys. Lett. \textbf{75}, 841 (2006);
Phys. Rev. E \textbf{75}, 022101 (2007); M. Perc and A. Szolnoki,
Phys. Rev. E \textbf{77}, 011904 (2008).

\bibitem{nowak:noise}
M. A. Nowak, S. Bonhoeffer, and R. M. May, Internat. J. Bifur.
Chaos \textbf{4} (1994), 3.

\bibitem{Tanimote2007pre}%noise in payoff
J. Tanimoto, Phys. Rev. E \textbf{76}, 041130 (2007).

\bibitem{Girvan2002pnas}
M. Girvan, M.~E.~J. Newman, Proc. Natl. Acad. Sci. U.S.A.
\textbf{99}, 7821 (2002).

\bibitem{Castellano2005aip}%voter and invasion dyanmics
C. Castellano, AIP Conf. Proc. \textbf{779}, 114 (2005).
\end{thebibliography}

\end{document}